\documentclass[sigconf]{acmart}

\usepackage{enumitem}

\title{\textsc{ThreatLens}: Evidence-Guided Ranking of High-Priority CVEs}

\author{Soroush Motamedi Sedeh}
\affiliation{%
  \institution{Simon Fraser University}
  \city{Burnaby}
  \state{British Columbia}
  \country{Canada}}
\email{sma266@sfu.ca}

\author{Panteha Shahrivar}
\affiliation{%
  \institution{Simon Fraser University}
  \city{Burnaby}
  \state{British Columbia}
  \country{Canada}}
\email{panteha_shahrivar@sfu.ca}

\author{Malaika Qureshi}
\affiliation{%
  \institution{Simon Fraser University}
  \city{Burnaby}
  \state{British Columbia}
  \country{Canada}}
\email{malaika_qureshi@sfu.ca}

\author{Ali Devjiani}
\affiliation{%
  \institution{Simon Fraser University}
  \city{Burnaby}
  \state{British Columbia}
  \country{Canada}}
\email{ali_devjiani@sfu.ca}

\author{Mohammad A. Tayebi}
\affiliation{%
  \institution{Simon Fraser University}
  \city{Burnaby}
  \state{British Columbia}
  \country{Canada}}
\email{tayebi@sfu.ca}

\copyrightyear{2026}
\acmYear{2026}
\setcopyright{cc}
\setcctype{by}
\acmConference[CIKM '26]{Proceedings of the 35th ACM International Conference on Information and Knowledge Management}{November 07--11, 2026}{Rome, Italy}
\acmBooktitle{Proceedings of the 35th ACM International Conference on Information and Knowledge Management (CIKM '26), November 07--11, 2026, Rome, Italy}
\acmDOI{10.1145/3799682.3840103}
\acmISBN{979-8-4007-2539-5/2026/11}

\begin{document}

\begin{abstract}

Security teams must prioritize vulnerabilities before exploitation evidence is complete. Existing signals, such as CVSS, EPSS, advisories, and public exploits, are useful but fragmented and time-sensitive; retrospective rankings can therefore overstate performance by using evidence unavailable at decision time. We present \textsc{ThreatLens}, a simple yet effective and deployment-realistic framework for CVE prioritization. \textsc{\textsc{ThreatLens}} ranks vulnerabilities at each review point using only cutoff-valid evidence and learns from future CISA KEV entries as weak supervision for exploitation relevance. Under forward-in-time, CVE-disjoint evaluation, \textsc{\textsc{ThreatLens}} significantly outperforms CVSS, EPSS, and rule-based evidence-fusion baselines. On the held-out test split, {\sc ThreatLens} surfaces 80.0\% of future KEV CVEs in the top 20, over three times EPSS at the same budget, and reaches 95.9\% in the top 50. Early-warning analysis further shows that \textsc{ThreatLens} identifies a substantial fraction of subsequent KEV entries before formal catalog inclusion, supporting timely, evidence-grounded triage.

\end{abstract}

\begin{CCSXML}
<ccs2012>
   <concept>
       <concept_id>10002978.10003006.10011634</concept_id>
       <concept_desc>Security and privacy~Vulnerability management</concept_desc>
       <concept_significance>500</concept_significance>
       </concept>
   
 </ccs2012>
\end{CCSXML}

\ccsdesc[500]{Security and privacy~Vulnerability management}

\keywords{Cyber Threat Intelligence, Vulnerability Prioritization, CVE Ranking, Exploitation Prediction}

\maketitle

\section{Introduction}

Modern vulnerability management is constrained by scale. New Common Vulnerabilities and Exposures (CVE) \cite{cve} are disclosed continuously, while security teams have limited time to investigate, validate, and remediate them. In this setting, treating every published vulnerability as equally urgent is impractical. At the same time, relying only on static severity scores such as the Common Vulnerability Scoring System (CVSS) is insufficient~\cite{cvss, chen2019twitter, sabottke2015twitter, nappa2015clones}: operational risk depends not only on technical severity, but also on exploitation likelihood, public exploit availability, advisory activity, affected product context, and how these signals evolve over time.

This creates a prioritization problem for threat intelligence analysts: given the evidence available today, which CVEs should be reviewed first? This problem is naturally a ranking task rather than a binary classification task \cite{liu2009ltr, burges2005ranknet}. Analysts rarely need an abstract prediction over all vulnerabilities; instead, they work under limited review budgets, such as inspecting the top 10, 20, or 50 CVEs in a triage cycle. A useful system should therefore rank vulnerabilities within time-bounded candidate pools using only evidence that would have been available at the decision point.

We present \textsc{ThreatLens}, an effective evidence-guided framework for time-aware CVE prioritization. \textsc{ThreatLens} models each vulnerability as a \emph{CVE-time snapshot}: a CVE observed at a cutoff date and represented only by evidence available up to that cutoff. The framework learns to rank vulnerabilities that later enter the CISA Known Exploited Vulnerabilities (KEV) catalog~\cite{kev, bod2201} above lower-priority candidates. We use KEV as weak supervision because it is public, timestamped, and tied to real-world exploitation~\cite{kev, bod2201, parla2024}. However, \textsc{ThreatLens} is not designed as a narrow KEV-membership classifier. Rather, KEV provides a proxy for exploitation-relevant prioritization under realistic time constraints.

\textsc{ThreatLens} integrates public vulnerability and threat- \linebreak intelligence evidence from multiple sources, including the National Vulnerability Database (NVD) vulnerability descriptions \cite{nvd}, CVSS metadata, Exploit Prediction Scoring System (EPSS) scores \cite{epssWebsite}, GitHub Security Advisory (GHSA) activity \cite{ghsa}, and public exploit signals~\cite{jacobs2021epss, jacobs2023epssv3, bozorgi2010, suciu2022ee, xiao2018riskprofiles}. To avoid temporal leakage, all evidence is aligned to the weekly cutoff at which the ranking decision is made. This design allows us to evaluate prioritization as it would occur in deployment: models are trained on past vulnerabilities, tested on future vulnerabilities, and assessed by whether they place operationally relevant CVEs near the top of weekly review queues.

% \begin{figure}[t]
%     \centering
%    \includegraphics[width=\columnwidth,height=90mm,keepaspectratio]{graphs/figure1.pdf}
%     \caption{Forward-looking vulnerability prioritization. At review time \(t\),
% analysts observe a pool of candidate CVEs and cutoff-valid public evidence.
% \textsc{\textsc{ThreatLens}} scores and ranks these candidates into a top-\(K\) review
% queue, aiming to surface vulnerabilities that later become operationally
% important, including those that enter KEV, before their importance is formally
% recognized.}
%     \label{fig:forward-prioritization}
% \end{figure}

We evaluate \textsc{ThreatLens} on a weekly large CVE snapshot dataset constructed from public vulnerability and threat-intelligence sources. Each snapshot captures the evidence available for a CVE at a specific cutoff, including NVD/CVSS information, EPSS scores, advisory evidence, and public exploit signals. Our deployment-style evaluation splits data forward in time, keeps CVEs disjoint across train, validation, and test sets, and ranks candidates within weekly review pools. On the held-out test split, where future-KEV CVEs account for only 2.50\% of samples, learned rankers substantially improve prioritization over non-learned baselines. {\sc ThreatLens} achieves 80.0\% Recall@20, more than tripling that of EPSS, as well as reaching 95.9\% Recall@50, and 70.8\% global PR-AUC, while early-warning analysis shows that learned scorers can surface KEV-bound vulnerabilities before formal KEV entry.

This paper makes four main contributions. First, we formulate CVE prioritization as a time-aware ranking problem over weekly analyst review queues. Second, we construct cutoff-aligned CVE snapshots that capture evolving public evidence while mitigating temporal leakage. Third, we develop \textsc{ThreatLens}\footnote{Source: https://github.com/smo165/ThreatLens} as an evidence-guided framework for CVE prioritization, with multiple rankers over structured and textual representations of cutoff-valid evidence. Fourth, we evaluate \textsc{ThreatLens} on a large-scale, forward-in-time, CVE-disjoint test set and show substantial improvements in top-K prioritization and early warning over standard baselines.

\begin{figure}[t]
\centering
\includegraphics[width=\columnwidth]{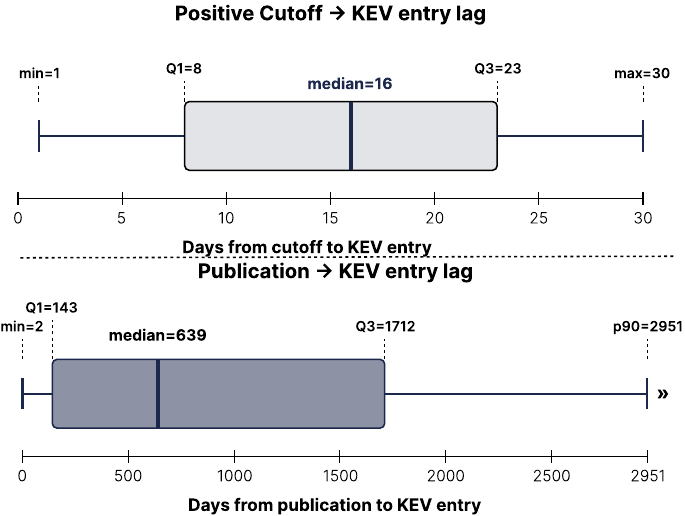}
\caption{KEV timing distributions. Publication-to-KEV lag is long-tailed
(median: 639 days), while positive cutoff-to-KEV lag is concentrated within
the 30-day horizon (median: 16 days). Q1/Q3 denote the 25th/75th percentiles,
and p90 the 90th percentile.}
%\caption{KEV timing distributions. Publication-to-KEV lag is highly
% long-tailed, with a median of 639 days, showing that many vulnerabilities
% enter KEV long after disclosure. In contrast, positive cutoff-to-KEV lag is
% concentrated within the 30-day supervision horizon, with a median of 16 days. Q1/Q3 denote
% the 25th/75th percentiles, and p90 denotes the 90th percentile, used instead
% of the maximum because of extreme outliers.}
\label{fig:KEV_lag}
\end{figure}

\section{Related Work}
Vulnerability prioritization requires more than static severity scoring. CVSS-based filtering often produces excessive alerts and can poorly reflect real-world exploitation \cite{cvss,allodi2014}, while EPSS and KEV provide complementary but incomplete exploitation signals \cite{epssWebsite,kev}. Exploitation may also precede CVSS publication, making temporally valid evaluation essential \cite{chen2019twitter,sabottke2015twitter,nappa2015clones}. Prior exploit-prediction work further identifies severe class imbalance and temporal leakage as key challenges \cite{bullough2017,xiao2018riskprofiles}.

EPSS introduced an open probabilistic framework for estimating exploitation likelihood \cite{jacobs2021epss}, with EPSS v3 combining severity, exploit availability, and advisory activity to substantially reduce remediation workload \cite{jacobs2023epssv3}. KEV has also been used as timestamped evidence of observed exploitation \cite{parla2024}, while recent work combines KEV, EPSS, and CVSS through sequential filtering \cite{shimizu2026}. Other studies show that learned models using vulnerability metadata, exploit evidence, timestamps, and references outperform static heuristics \cite{bozorgi2010,suciu2022ee}. These results suggest that EPSS, KEV, exploit availability, and advisory activity provide complementary signals, but have not been systematically integrated within a unified, temporally controlled ranking framework.

Temporal realism is particularly important because CVE evidence accumulates after disclosure. Prior work highlights leakage and the ``perfect labeling assumption,'' where models are evaluated using information unavailable at prediction time \cite{le2022survey,iannone2024}. \textsc{ThreatLens} addresses this by constructing cutoff-aligned weekly snapshots, using only evidence available by each cutoff, enforcing forward-in-time splits, and preventing CVE overlap across splits.

More broadly, prior work suggests that vulnerability prioritization should be evaluated as an operational decision process rather than retrospective exploit classification. \textsc{ThreatLens} builds on this perspective by treating public evidence as time-bounded input to a recurring analyst review queue. It ranks CVE-time snapshots using only cutoff-valid evidence and measures whether future KEV entries are surfaced within small weekly top-$K$ budgets, reflecting decisions under evolving evidence and limited triage capacity.

\section{Data Characteristics}

\textbf{Data sources and snapshot construction.} \textsc{ThreatLens} integrates public vulnerability and threat-intelligence sources: NVD descriptions and metadata~\cite{nvd}, CISA KEV as a weak supervision signal~\cite{kev}, EPSS for time-varying exploitation likelihood~\cite{epssWebsite}, GHSA for advisory context~\cite{ghsa}, as well as Exploit-DB~\cite{exploitdb} and Metasploit~\cite{metasploit} for public exploit availability. We convert these sources into weekly CVE-time snapshots. At each cutoff, the eligible pool includes CVEs published by that date that have not yet entered KEV. {\it Positives} are CVEs that enter KEV within the following 30 days; {\it negatives} are sampled from the remaining eligible CVEs at the same cutoff.

The final corpus contains 269,331 snapshots across 233 weekly cutoffs from October 2021 to March 2026, covering 158,217 unique CVEs. Positives account for 1.96\% of snapshots, and 45.7\% of CVEs appear at multiple cutoffs. This design matches the operational setting: each week, a model must rank not-yet-confirmed vulnerabilities using only evidence available up to that cutoff. KEV is useful for this setting because inclusion requires a CVE ID, reliable evidence of active exploitation in the wild, and a clear remediation action. Thus, entry into KEV is not merely a severity label; it is a timestamped operational signal of confirmed exploitation relevance and remediation urgency.

\noindent \textbf{Temporal imbalance and evidence dynamics.} Each cutoff defines a separate ranking pool. The median weekly pool contains 510 candidate CVEs, including 10 positives and 500 negatives; the mean pool contains 1,155.9 CVEs, including 22.7 positives. This imbalance reflects the deployment scenario, where analysts must select a small review set from a large weekly candidate pool. Negative sampling keeps enrichment, training, and evaluation tractable while preserving same-week ranking structure.

Repeated snapshots are common but not redundant. Overall, 72,374 CVEs appear
at multiple cutoffs, yielding 111,114 consecutive same-CVE snapshot pairs.
Across these transitions, cutoff-valid evidence changes often: EPSS score,
EPSS percentile, and GHSA advisory count change in 60.9\%, 87.5\%, and
28.4\% of pairs, while public exploit indicators rarely change. These dynamics
motivate modeling CVEs as time-indexed instances rather than static records.

\noindent \textbf{KEV timing.} We distinguish KEV membership from KEV timing. KEV membership serves as a weak proxy for confirmed exploitation relevance, while KEV timing serves as a weak proxy for operational urgency. As shown in Figure~\ref{fig:KEV_lag}, publication-to-KEV lag is long-tailed, with a median of 639 days. By construction, positive cutoff-to-KEV lag is concentrated within the 30-day supervision window, with a median of 16 days. KEV timing should therefore not be interpreted as exploitation onset; rather, it reflects exploitation, disclosure, cataloging, and reporting dynamics.

\begin{figure}[t]
\centering
\includegraphics[width=\columnwidth]{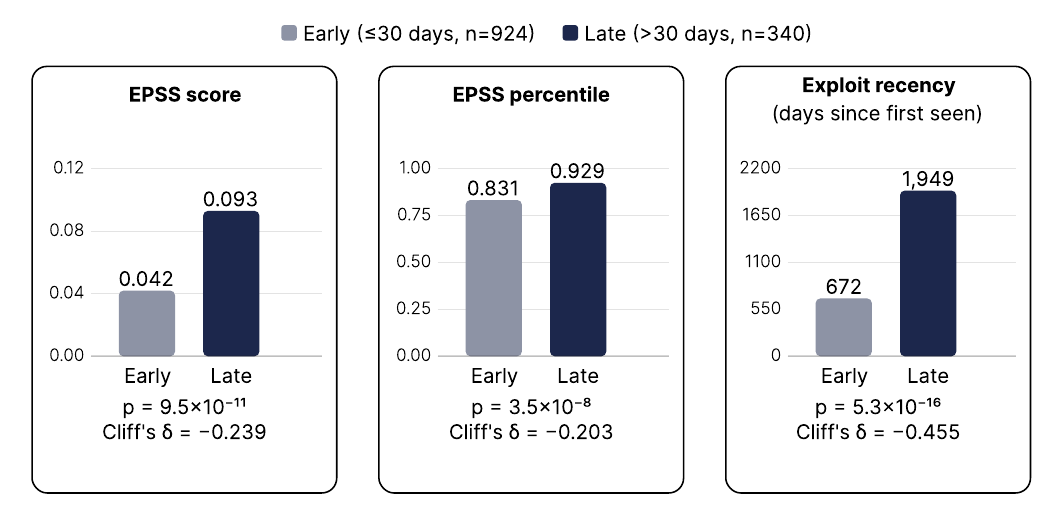}
\caption{Pre-entry evidence differences between early and late KEV entrants.
Late entrants show higher EPSS scores and EPSS percentiles, while early
entrants show shorter exploit recency, measured as days since first public
exploit evidence. All three differences are statistically significant.}

\label{fig:pre-entry_evidence_distribution}
\end{figure}

To assess whether KEV timing aligns with observable pre-entry evidence, we partition KEV CVEs by first-eligible-cutoff lead time. The first-eligible cutoff is the earliest weekly cutoff at which a CVE enters the eligible pool. We define \emph{early} KEV CVEs as those entering KEV within 30 days of this cutoff, and \emph{late} KEV CVEs as the remainder. As shown in Figure~\ref{fig:pre-entry_evidence_distribution}, Mann--Whitney tests ~\cite{mannwhitney} on the last pre-KEV snapshots show significant differences in EPSS score, EPSS percentile, and public exploit recency, while GHSA advisory count is not significantly different (\(p=0.499\)). These results suggest that early and late KEV trajectories differ under cutoff-valid evidence. However, the lower EPSS medians for early-group CVEs also indicate that KEV timing captures operational and cataloging factors not reducible to EPSS magnitude alone.

% To test whether KEV timing corresponds to measurable evidence differences before entry, we partition KEV CVEs by first-eligible-cutoff lead time (where first-eligible cutoff is the earliest weekly cutoff at which the CVE appears in the eligible pool): early if first-eligible-cutoff-to-KEV $\leq 30$ days and late otherwise. Mann–Whitney tests on last pre-KEV snapshots reveal significant distributional differences in EPSS score, EPSS percentile, and public exploit recency (see Figure 3), while GHSA advisory count is not significantly different ($p=0.499$). These results are associational rather than causal and indicate that early and late KEV trajectories are not distributionally identical in cutoff-valid evidence. The lower EPSS medians for early-group CVEs suggest that KEV timing may capture operational and catalog dynamics not reducible to EPSS magnitude alone.

\section{Vulnerability Prioritization}
\label{sec:methodology}

\begin{figure}[t]
    \centering
      \includegraphics[width=\columnwidth]{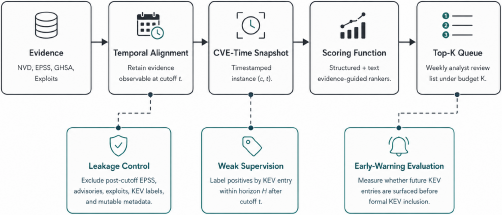}
    \caption{ThreatLens workflow. Weekly CVE snapshots are constructed from time-bounded evidence, ranked under fixed analyst review budgets, and evaluated for pre-KEV early warning.}
    \label{fig:threatlens-workflow}
\end{figure}

\textsc{ThreatLens} is a cutoff-aligned CVE prioritization pipeline that
converts heterogeneous public vulnerability intelligence into weekly ranked
review queues. As shown in Figure~\ref{fig:threatlens-workflow}, the pipeline
constructs time-bounded CVE snapshots at each weekly cutoff, ranks candidate
vulnerabilities under fixed analyst review budgets, and evaluates whether
high-risk CVEs are surfaced before KEV entry. Its main methodological
contribution is not a new model architecture, but a deployment-style
formulation and data-construction process that preserves temporal validity,
controls leakage, and supports fair comparison across scoring methods.
\subsection{Problem Formulation}

We formulate vulnerability prioritization as a periodic ranking problem. Let $\mathcal{T}$ be a sequence of regularly spaced review times. At each cutoff $t \in \mathcal{T}$, the system observes a candidate pool $\mathcal{C}_t$ of CVEs that are known by time $t$ and have not already entered KEV. For each CVE $c \in \mathcal{C}_t$, let $x_{c,t}$ denote the evidence available at cutoff $t$. \textsc{ThreatLens} learns a scoring function

\[
s_\theta(c,t) = f_\theta(x_{c,t}),
\]

where larger scores indicate higher priority for analyst review. The output is a ranked list

\[
\pi_t = \operatorname{sort}_{c \in \mathcal{C}_t}(s_\theta(c,t)),
\]

from which analysts inspect the top-$K$ CVEs according to their review budget.

We use future entry into the KEV catalog as weak supervision. Let $T(c)$ be the KEV entry date for CVE $c$, and let $H$ be the prediction horizon. A snapshot $(c,t)$ is positive if the CVE enters KEV after the cutoff and within the horizon:

\[
y^{(H)}_{c,t} =
\begin{cases}
1, & \text{if } t < T(c) \leq t + H, \\
0, & \text{otherwise.}
\end{cases}
\]

This label is time-dependent: the same CVE may be negative at earlier cutoffs and positive at later cutoffs. Note that KEV is treated as a weak operational proxy for exploitation relevance, not as complete ground truth. This proxy is useful because it provides
a public, timestamped signal that allows the model to learn patterns
associated with vulnerabilities that become operationally important sooner than others.

\subsection{Snapshot Construction}

\textsc{ThreatLens} represents each example as a CVE-time snapshot $(c,t)$. For each cutoff, we construct the eligible candidate pool as

\[
\mathcal{C}_t =
\{c : \operatorname{pub}(c) \leq t \land (T(c) \text{ undefined} \lor T(c) > t)\}.
\]

We retain all positive snapshots and sample negative snapshots from the remaining eligible CVEs at the same cutoff. This preserves the cutoff-specific ranking structure: future-KEV CVEs compete against other CVEs that would have been visible to analysts at the same review time. Negative sampling keeps enrichment, training, and evaluation computationally feasible while maintaining strong class imbalance.

For each selected snapshot, \textsc{ThreatLens} constructs a cutoff-valid evidence representation:
\[
x_{c,t} = \operatorname{Evidence}(c,\leq t).
\]
Evidence is drawn from NVD vulnerability context, CVSS information, EPSS, GHSA advisories, Exploit-DB, and Metasploit. Time-varying sources are aligned to the cutoff: EPSS uses the cutoff date or the nearest prior available date; advisories are filtered by publication or modification time; Exploit-DB entries are filtered by exploit publication date; and Metasploit evidence uses disclosure date as a proxy for public availability.

\subsection{Temporal Leakage Mitigation}

Because the task is temporal, leakage control is central to the methodology. \textsc{ThreatLens} removes or transforms evidence that could reveal information unavailable at ranking time. Raw CVE and advisory identifiers are scrubbed from model-facing text to reduce memorization. Exact cutoff dates, advisory dates, EPSS-as-of dates, and exploit publication dates are excluded; when temporal context is useful, it is encoded through relative or aggregate features such as CVE age, advisory counts, exploit counts, EPSS score and percentile, and days since first public exploit evidence. 

NVD-derived fields are leakage-mitigated but not perfectly historical because we do not reconstruct exact historical NVD record revisions. We therefore remove mutable or leakage-prone NVD fields, including last-modified metadata, vulnerability status, references, URLs, tags, and raw metric-key metadata. We further test the effect of potentially non-historical NVD analysis fields through CVSS baselines and ablations that remove CVSS, Common Weakness Enumeration (CWE) \cite{cwe}, and Common Platform Enumeration (CPE) information \cite{cpe}.

\subsection{Model-Facing Representations}

\textsc{ThreatLens} derives two representations from each snapshot:

\noindent \textbf{Evidence text} serializes cutoff-valid vulnerability context, severity summaries, EPSS values, advisory evidence, public exploit evidence, and relative temporal information. This representation is used by the neural and TF-IDF text scorers.

\noindent \textbf{Structured features} encode cutoff-valid tabular signals, including
EPSS availability, score, and percentile; CVSS base score; advisory
availability and counts; public exploit availability and counts; CVE
age features; exploit recency features; and interaction features such as EPSS with public exploit evidence, EPSS with advisory evidence,
and CVSS with EPSS. This representation is used by the logistic regression model.

Using both representations allows us to test whether performance comes from cutoff-aligned evidence itself, compact structured summaries, or neural text representation learning.

\subsection{Scoring Models}

\textsc{ThreatLens} evaluates multiple scorers over the same CVE-time snapshots.

\noindent \textbf{Neural text ranker.}
The neural scorer encodes evidence text using BGE-large-en-v1.5 \cite{bge} and applies an MLP scoring head \cite{burges2005ranknet}. It is trained with same-review-time pairwise supervision: a positive snapshot $(c^+,t)$ and a negative snapshot $(c^-,t)$ are sampled from the same cutoff, and the model optimizes

\[
\mathcal{L}
=
-\log \sigma
\left(
s_\theta(c^+,t) - s_\theta(c^-,t)
\right).
\]

The model is trained to assign higher scores
to positives than to same-cutoff negatives. Rather than treating exploitation relevance only as a standalone
binary classification task, this objective learns relative priority
among CVEs that compete for analyst attention within the same
review-time pool, matching the deployment setting.

\noindent \textbf{Structured scorer.}
We train a class-weighted logistic regression over structured features. Features include EPSS score and percentile, CVSS
base score, advisory availability and counts, public exploit availability
and counts, CVE age, exploit recency, and interaction features. Predicted probabilities are used as priority scores. This scorer tests how effectively explicit cutoff-valid tabular summaries of the enriched evidence can support prioritization.

\noindent \textbf{Shallow text scorer.}
We train a class-weighted TF-IDF logistic regression model \cite{salton1988term,tf-idf} over the same evidence text used by the neural ranker. Predicted probabilities are used as ranking scores. This tests whether neural gains exceed shallow lexical matching.

% Additional details for these scorers are given in section 5.2.

\section{Evaluation}
\label{sec:evaluation}

We evaluate \textsc{ThreatLens} under a deployment-style vulnerability
prioritization setting. At each weekly review point, the system
observes only evidence available up to that cutoff and ranks
candidate CVEs for analyst review. The evaluation is designed to
answer four research questions:

\begin{itemize}[label=$\diamond$]
\item {\bf RQ1:} How well do learned evidence-guided scorers prioritize
future KEV CVEs under fixed analyst review budgets?

\item {\bf RQ2:} Do learned scorers outperform standalone signals such as
CVSS, EPSS, and public exploit availability?

\item {\bf RQ3:} Can \textsc{ThreatLens} surface future KEV CVEs before their KEV
entry date?

\item \textbf{RQ4:} Are the gains driven by a single shortcut signal, such as
EPSS or CVSS, or by multi-source evidence integration?

\end{itemize}

\subsection{Experimental Design}
\label{subsec:experimental-setup}
We instantiate \textsc{ThreatLens} on weekly CVE snapshots with a 30-day
horizon. At each cutoff \(t\), the candidate pool includes CVEs published by
\(t\) not yet in CISA KEV. A snapshot is positive if the CVE enters KEV within
30 days, and negative otherwise. For each cutoff, we retain all positives and sample negatives from the
same eligible pool at a 50:1 ratio, preserving the same-week ranking
setting faced by analysts. To avoid temporal leakage, we use a
forward-in-time split and enforce CVE-disjointness by removing from later
splits any CVE that appeared earlier. Thus, validation and test are
separated from training both by time and by vulnerability identity.
Table~\ref{tab:split} summarizes the final split; the strict test set
contains 11,779 snapshots, 11,193 unique CVEs, and 295 positives
(2.50\%).

\begin{table}[t]

\caption{Temporal and CVE-disjoint split for CVE snapshots across cutoffs from October 2021 to March 2026.}
\label{tab:split}
\centering

\begin{tabular}{lrrrr}
\toprule
Split & Snapshots & Unique CVEs & Positives & Pos. Rate \\
\midrule
Train & 226,593 & 136,435 & 4,443 & 1.96\% \\
Validation & 11,147 & 10,589 & 307 & 2.75\% \\
Test & 11,779 & 11,193 & 295 & 2.50\% \\
\bottomrule
\end{tabular}
\end{table}

\subsubsection{Baselines and ThreatLens Variants}
\label{subsec:models-baselines}

We compare learned \linebreak\textsc{ThreatLens} scorers against non-learned
prioritization baselines. All scorers operate on the same cutoff-aligned
snapshot dataset and are evaluated using the same weekly ranking
pools.

\begin{itemize}[label=$\triangleright$]

\item  \textbf{\textsc{CVSS}.}
This baseline ranks candidate CVEs by their CVSS base score, representing severity-based vulnerability prioritization.

\item  \textbf{\textsc{EPSS}.} This baseline ranks candidate CVEs by their cutoff-valid EPSS score, representing exploitation-likelihood-based vulnerability prioritization.

\item \textbf{\textsc{Expert-Rule}.}
This baseline applies a fixed scoring rule over cutoff-valid
structured evidence, including EPSS, CVSS, advisory activity, and public
exploit signals. It serves as a simple evidence-fusion baseline, testing
whether learned rankers provide value beyond manually combining the same
high-level signals.

\item \textbf{\textsc{TL-NR}.}
\textsc{TL-NR} is the neural text-ranking variant of \linebreak \textsc{ThreatLens}.
It encodes cutoff-valid evidence text using BGE-large-en-v1.5 with
CLS pooling, followed by a two-layer MLP scoring head. The encoder and head
are fine-tuned jointly using same-cutoff pairwise supervision: each pair
contains one positive and one negative snapshot sampled from the same weekly
candidate pool. This objective trains the model to assign higher priority
scores to future KEV snapshots than to competing same-week candidates.

% \item  \textbf{\textsc{TL-NR}.} This is a variation of {\sc ThreatLens} which uses a neural text ranker. {\sc TL-NR}
% The \textsc{ThreatLens} uses BGE-large-en-v1.5 with
% CLS pooling and applies a two-hidden-layer MLP scoring head with hidden dimension 256 and dropout 0.1. Evidence
% text is tokenized with maximum sequence length 512, and the
% encoder and scoring head are fine-tuned jointly. The model is trained
% with same-week pairwise supervision: for each training pair, a positive
% snapshot and a negative snapshot are sampled from the same weekly
% cutoff.

% The neural ranker is trained for 10 epochs with 50,000 same-week pairs
% per epoch, pair batch size 64, and evaluation batch size 256. We use
% AdamW with encoder learning rate $10^{-5}$, head learning rate
% $10^{-4}$, weight decay $10^{-4}$, mixed-precision training, and
% gradient clipping at 1.0.

% \item \textbf{\textsc{TL-HGB}} is a variation of \textsc{ ThreatLens} that uses median imputation followed by a histogram gradient boosting classifier trained with class-balanced sample weights. % % 

\item \textbf{\textsc{TL-LR}.}
\textsc{TL-LR} is a structured \textsc{ThreatLens} variant based on
logistic regression. It uses cutoff-valid tabular features with median
imputation, feature standardization, and class-weighted training.

\item \textbf{\textsc{TL-TF-IDF}.}
\textsc{TL-TF-IDF} is a shallow text-ranking variant of \linebreak\textsc{ThreatLens}.
It consumes the same evidence text as \textsc{TL-NR}, represents each
snapshot using word unigram and bigram TF-IDF features with sublinear
term-frequency scaling, and trains a class-weighted logistic regression
scorer.

% \item \textbf{\textsc{Structured Rule}.} This baseline combines structured signals such
% as EPSS, CVSS, advisory counts, and public exploit evidence using
% a hand-weighted rule score. It tests whether simple rule-based
% evidence fusion can match learned scoring.

% \item  \textbf{Non-learned baselines.} We compare against CVSS, EPSS at the cutoff, public exploit count, and a simple hand-designed evidence-fusion rule. The rule assigns each CVE a priority score by starting from its EPSS value, adding a normalized contribution from CVSS severity, and then increasing the score when public exploits or security advisories are available. These baselines test whether individual risk signals or fixed expert rules are sufficient for vulnerability prioritization.

\end{itemize}

% We compare against four reference baselines: CVSS base score, EPSS
% score at the cutoff, public exploit count, and a fixed structured rule
% combining EPSS, CVSS, public exploit availability, and advisory
% availability: 

% \[
% \begin{aligned}
% s_{\mathrm{rule}}
% &= \mathrm{EPSS}
% + 0.5 \cdot \frac{\mathrm{CVSS}}{10} \\
% &\quad
% + 0.75 \cdot \mathbb{1}[\mathrm{public\_exploit\_count} > 0] \\
% &\quad
% + 0.25 \cdot \mathbb{1}[\mathrm{advisory\_count} > 0].
% \end{aligned}
% \]

% These baselines test whether individual risk signals or
% simple hand-designed evidence fusion are sufficient for the
% prioritization task.

\subsubsection{Evaluation Metrics}
For top-\(K\) evaluation, snapshots are \linebreak grouped by cutoff date and ranked
within the corresponding weekly candidate pool. We use Recall@\(K\) as
the primary metric because analysts operate under limited review
budgets. We report Recall@20 and Recall@50, computed as the total number of positives surfaced across weekly top-$K$ lists divided by the total number of positives across those weekly pools.

We also report global PR-AUC and macro PR-AUC by cutoff, since positive
snapshots are rare \cite{davis2006pr, saito2015pr}. All methods use the
same temporal split, CVE-disjoint filtering, candidate pools, and
metrics, so performance differences reflect scoring quality rather than
evaluation artifacts.

% At evaluation time, each scorer assigns a score to every snapshot in the
% validation or test split. For top-$K$ evaluation, snapshots are grouped by
% cutoff date and ranked within the corresponding weekly candidate pool.
% Thus, the main ranking metrics evaluate same-week prioritization within
% constructed cutoff-specific pools.

% We use Recall@$K$ as the primary metric because the operational
% question is whether future KEV CVEs appear within a limited analyst
% review budget. We report Recall@20 and Recall@50, corresponding to
% review budgets of 20 and 50 CVEs per weekly cycle. We report both
% macro Recall@$K$, which averages recall across cutoffs, and micro
% Recall@$K$, which aggregates captured positives across all cutoffs. We
% also report global PR-AUC and macro PR-AUC by cutoff as complementary
% metrics. PR-AUC is especially appropriate because positive snapshots
% are rare \cite{davis2006pr, saito2015pr}.

% All learned scorers and baselines are evaluated under the same temporal
% split, CVE-disjoint filtering, weekly candidate pools, and ranking metrics.
% Therefore, performance differences reflect scoring quality rather than
% differences in candidate-pool construction or evaluation protocol.

\subsection{Main Ranking Results}
\label{subsec:main-ranking-results}

Table~\ref{tab:scorer-comparison} reports held-out test performance. The
non-learned baselines are substantially weaker than the learned scorers.
EPSS is the strongest standalone baseline 
while CVSS achieves the highest non-learned Recall@50. However,
neither EPSS nor CVSS approaches the performance of learned evidence
fusion. This confirms that individual risk signals are useful but
insufficient for deployment-style prioritization. At a weekly review budget of 20 CVEs, \textsc{TL-LR} surfaces more than three times as many future KEV vulnerabilities as EPSS (80.0\% vs 24.7\% Recall@20).

\begin{table}[t]
\caption{Held-out test performance of baselines and \textsc{ThreatLens} variants.}
\label{tab:scorer-comparison}
\centering
\begin{tabular}{lrrrr}
\toprule
Method & R@20 & R@50 & PR-AUC & MPR-AUC \\
\midrule
CVSS & 19.3\% & 42.4\% & 6.5\% & 9.4\% \\
% Exploit count & 17.6\% & 33.6\% & 5.2\% & 8.0\% \\
EPSS & 24.7\% & 36.9\% & 8.0\% & 12.0\% \\
Expert-Rule & 19.0\% & 26.4\% & 4.5\% & 8.3\% \\
\textsc{TL-NR} & 75.6\% & 92.2\% & 57.1\% & 54.6\% \\
% TL-Structured HGB & 68.8\% & 93.9\% & 56.6\% & 60.7\% \\
\textsc{TL-LR} & \textbf{80.0\%} & \textbf{95.9\%} & \textbf{70.8\%} & \textbf{73.9\%} \\
\textsc{TL-TF-IDF} & 61.7\% & 76.3\% & 43.8\% & 51.5\% \\
\bottomrule
\end{tabular}
\end{table}

The learned scorers provide large gains. \textsc{TL-TF-IDF} substantially
outperforms the non-learned baselines, indicating that the evidence text
contains useful prioritization signal. \textsc{TL-NR} further
improves over \textsc{TL-TF-IDF}, increasing Recall@20 from 61.7\% to
75.6\%, Recall@50 from 76.3\% to 92.2\%, and global PR-AUC from
43.8\% to 57.1\%. This shows that fine-tuned neural ranking captures
useful semantic signal beyond shallow lexical matching.

The strongest aggregate scorer is \textsc{TL-LR}. It
achieves 80.0\% Recall@20, 95.9\% Recall@50, 70.8\%
global PR-AUC, and 73.9\% macro PR-AUC. This result is notable because
the model is lightweight and interpretable, yet highly effective when
trained on carefully constructed cutoff-valid evidence. It suggests that
much of the prioritization signal is captured by the interaction of
exploitation likelihood, severity, advisory activity, public exploit evidence,
and temporal features. The largest positive standardized coefficients of \textsc{TL-LR} correspond to GHSA advisory count, public exploit count, EPSS score, CVSS base score, EPSS availability, and exploit-timing features, indicating that the strongest scorer combines multiple evidence sources rather than relying on one isolated signal.

A notable result is that the strongest model is not the largest neural model, but a structured logistic ranker. This does not weaken the case for {\sc ThreatLens}; instead, it supports the central thesis of the paper. Once evidence is temporally aligned and leakage is controlled, compact signals such as EPSS movement, advisory activity, exploit availability, CVSS severity, and recency encode substantial prioritization value. The gain comes less from model complexity than from constructing the right decision-time representation.

% Table~\ref{tab:structured-logreg-coefficients} reports the largest positive
% standardized coefficients of the structured logistic scorer. The strongest
% weights correspond to GHSA advisory count, public exploit count, EPSS
% score, CVSS base score, EPSS availability, and exploit-timing features.
% This indicates that the strongest scorer does not rely on one isolated
% signal. Instead, it combines multiple evidence sources that may accumulate
% before formal KEV cataloging.

% \begin{table}[t]
% \centering
% \caption{Largest positive standardized coefficients of the structured logistic scorer.}
% \label{tab:structured-logreg-coefficients}
% \begin{tabular}{lr}
% \toprule
% Feature & Coefficient \\
% \midrule
% GHSA advisory count & 2.356 \\
% Log public exploit count & 1.887 \\
% EPSS score & 0.884 \\
% CVSS base score & 0.534 \\
% EPSS available & 0.399 \\
% Log days since first public exploit & 0.390 \\
% EPSS percentile & 0.235 \\
% \bottomrule
% \end{tabular}
% \end{table}

\subsection{Early-Warning Evaluation}

The main ranking evaluation measures whether \textsc{ThreatLens} surfaces
future KEV CVEs within the 30-day supervision horizon. We also evaluate
whether models can surface KEV CVEs before their KEV entry date,
regardless of that training horizon.

For each held-out KEV CVE \(v\) and review budget \(K\), we identify the
earliest weekly cutoff \(t\) before KEV entry at which \(v\) appears in the
top-\(K\) ranked list. We define the resulting lead time as:

\[
\mathrm{LeadTime}(v)
=
t_{\mathrm{KEV}}(v)
-
\min \left\{
t \;:\; v \in \mathrm{Top}\text{-}K(t),\; t < t_{\mathrm{KEV}}(v)
\right\}
\]

\noindent where \(t_{\mathrm{KEV}}(v)\) denotes the KEV entry date of \(v\). We evaluate early-warning performance under two settings. The
\textit{strict forward-time setting} uses the same held-out test split as
the main ranking evaluation and contains 92 KEV CVEs with pre-KEV
snapshots. This setting provides the most realistic estimate of future
deployment performance. The \textit{10-fold} KEV-target setting evaluates
broader coverage over 1,264 KEV CVEs with pre-KEV snapshots. Each fold
holds out a disjoint subset of KEV CVEs as early-warning targets. This
setting relaxes strict chronological ordering but provides broader
historical coverage.

In both settings, target CVEs are ranked against the constructed
background candidate pool at each cutoff. Training and validation CVEs
may appear as background candidates in the weekly ranking lists, but
early-warning hits are counted only for held-out target CVEs. This reflects
a deployment-style review scenario in which a future KEV CVE must
compete against a broader set of vulnerabilities known at the same time.

\begin{table}[t]
\caption{Early-warning performance before KEV entry. \textit{Forward}
denotes the strict forward-time held-out test setting; \textit{10-fold}
denotes CVE-disjoint KEV-target folds for broader historical coverage.
Top-$K$ values report the percentage of held-out KEV CVEs surfaced before
KEV entry. Med@50 reports the median top-50 warning time in days.}
\label{tab:early-warning-comparison}
\centering
\footnotesize
\setlength{\tabcolsep}{2.2pt}
\renewcommand{\arraystretch}{1.04}
\resizebox{0.99\columnwidth}{!}{%
\begin{tabular}{llrrrr}
\toprule
Setting & Method & Top 10 & Top 20 & Top 50 & Med@50 \\
\midrule
Forward & CVSS & 0.0\% & 5.4\% & 25.0\% & 14 \\
% Forward & Exploit count & 3.3\% & 5.4\% & 15.2\% & 15.5 \\
Forward & EPSS & 10.9\% & 15.2\% & 27.2\% & 24 \\
Forward & Expert-Rule & 6.5\% & 9.8\% & 18.5\% & 24 \\
Forward & \textsc{TL-NR} & 52.2\% & 67.4\% & 89.1\% & 25 \\
% Forward & TL-Struct. HGB & 58.7\% & 75.0\% & 94.6\% & 25 \\
Forward & \textsc{TL-LR} & \textbf{67.4\%} & \textbf{84.8\%} & \textbf{97.8\%} & 25 \\
Forward & \textsc{TL-TF-IDF} & 52.2\% & 63.0\% & 78.3\% & 24 \\
\midrule
10-fold & CVSS & 1.0\% & 4.0\% & 15.3\% & 23 \\
% 10-fold & Exploit count & 6.9\% & 9.6\% & 18.6\% & 25 \\
10-fold & EPSS & 13.9\% & 20.2\% & 30.1\% & 28 \\
10-fold & Expert-Rule & 12.3\% & 16.5\% & 27.8\% & 28 \\
10-fold & \textsc{TL-NR} & 28.1\% & 44.1\% & 58.5\% & 28 \\
% 10-fold & TL-Struct. HGB & 32.4\% & 44.0\% & 59.3\% & 28 \\
10-fold & \textsc{TL-LR} & 32.5\% & 42.9\% & 55.5\% & 28 \\
10-fold & \textsc{TL-TF-IDF} & \textbf{33.1\%} & \textbf{46.0\%} & \textbf{60.7\%} & 28 \\
\bottomrule
\end{tabular}%
}
\end{table}
The results reported in Table~\ref{tab:early-warning-comparison} show that learned scorers substantially outperform standalone
baselines. In the strict forward-time setting, \textsc{TL-LR} performs
strongest across review budgets, surfacing 90 of 92 held-out KEV CVEs in
the top 50 before KEV entry. By contrast, EPSS surfaces only 25 of 92 at
the same budget. \textsc{TL-NR} also remains
strong, especially at larger review budgets. In the broader 10-fold KEV-target setting, \textsc{TL-TF-IDF} achieves the strongest
coverage across review budgets. \textsc{TL-LR} and
\textsc{TL-NR} remain competitive, particularly at top-20 and
top-50. This suggests that textual evidence is especially useful for broad
historical KEV-target coverage, while structured evidence generalizes
particularly well under the strict chronological setting.

Across settings, learned scorers also provide substantial warning time
before KEV entry. The strongest learned methods achieve median top-50
warning times of approximately 25 days in the forward-time setting and
28 days in the 10-fold KEV-target setting. Overall, these results show that
observable public evidence often accumulates before KEV cataloging and
that \textsc{ThreatLens} can exploit this evidence to surface vulnerabilities earlier
than standalone signals.

A comparison of the two settings, {\it strict-forward-time} and {\it 10-fold}, shows that forward-time evaluation yields higher recall because the recent held-out KEV cases contain structured signals that transfer well from training, such as EPSS movement, severity, advisory presence, and exploit metadata. In contrast, the 10-fold setting covers a larger and more heterogeneous historical KEV population, where pre-KEV evidence patterns are less consistent, reducing recall for all methods. However, \textsc{TL-TF-IDF} performs best in 10-fold, suggesting that lexical cues help recover recurring historical patterns that compact structured features may miss.

\subsection{Ablation Study}
\label{subsec:ablation}

We perform ablations to test whether \textsc{ThreatLens}' gains are primarily
explained by dominant signals such as EPSS and CVSS/NVD-derived analysis
fields, or whether performance persists when these signals are removed. All
ablations use the same rows, labels, cutoffs, temporal/CVE-disjoint split, and
evaluation protocol as the main experiments. For the neural text ranker, \textsc{TL-NR} w/o EPSS removes the temporal EPSS
block from the model-facing evidence text. \textsc{TL-NR} w/o NVD Analysis
removes CVSS, CWE, CPE, as well as CVSS score/vector fragments from NVD-derived text.
For the structured logistic model, \textsc{TL-LR} w/o EPSS removes EPSS-derived
features and EPSS interaction terms, while \textsc{TL-LR} w/o CVSS removes
CVSS-derived features and the CVSS--EPSS interaction term.

Table~\ref{tab:ablation} shows that EPSS contributes useful signal, but does
not explain \textsc{ThreatLens}' gains by itself. Removing EPSS reduces \textsc{TL-NR}'s PR-AUC from 57.1\% to 45.6\%, but the ablated model still
substantially outperforms the EPSS baseline. Similarly, \textsc{TL-LR} w/o
EPSS retains strong top-$K$ performance, achieving 77.3\% Recall@20 and
96.3\% Recall@50. Removing CVSS-derived information also preserves strong performance.
\textsc{TL-LR} w/o CVSS achieves 79.7\% Recall@20 and 94.2\% Recall@50. For \textsc{TL-NR}, removing NVD analysis fields does not degrade performance
and improves several aggregate metrics. This result is consistent with the
possibility that some NVD-derived analysis fields add noise or encourage
less robust textual shortcuts, while the remaining cutoff-valid evidence
remains sufficient for prioritization.

Overall, the ablations support the conclusion that \textsc{ThreatLens}
benefits from integrating multiple cutoff-aligned evidence sources rather
than simply reproducing EPSS, CVSS, or NVD-derived metadata. The ablation results also
motivate hybrid designs that combine the temporal robustness and
interpretability of structured features with the semantic coverage of text
representations.

\begin{table}[t]
\caption{Ablation results on the held-out test split. }
\label{tab:ablation}
\centering
\resizebox{\columnwidth}{!}{%
\begin{tabular}{lrrrr}
\toprule
Method & R@20 & R@50 & PR-AUC & MPR-AUC \\
\midrule
EPSS baseline & 24.7\% & 36.9\% & 8.0\% & 12.0\% \\
\midrule
\textsc{TL-NR} & 75.6\% & 92.2\% & 57.1\% & 54.6\% \\
\textsc{TL-NR} w/o EPSS & 62.7\% & 90.2\% & 45.6\% & 46.5\% \\
\textsc{TL-NR} w/o NVD Analysis & 76.6\% & 94.6\% & 61.9\% & 63.1\% \\
\midrule
\textsc{TL-LR} & 80.0\% & 95.9\% & 70.8\% & 73.9\% \\
\textsc{TL-LR} w/o EPSS & 77.3\% & 96.3\% & 65.2\% & 67.8\% \\
\textsc{TL-LR} w/o CVSS & 79.7\% & 94.2\% & 65.8\% & 69.6\% \\
\bottomrule
\end{tabular}%
}
\end{table}

\section{Operational Implications}
Our results suggest that \textsc{ThreatLens}' value comes primarily from cutoff-aligned evidence construction, rather than from a single dominant model family. Compact structured features generalize well over time because they summarize heterogeneous vulnerability evidence into stable risk signals, rather than relying on specific product names, vendor ecosystems, or repeated advisory language. Text-based models are complementary: product names, vulnerability classes, advisory phrases, exploit terminology, and recurring description patterns provide useful lexical and semantic anchors, especially for broad historical KEV coverage. \textsc{TL-TF-IDF} preserves sparse lexical cues directly, while \textsc{TL-NR} compresses evidence text into semantic representations. Together, these models provide complementary views of the same cutoff-valid evidence.

This complementarity has practical implications for analyst-facing deployment. Structured scoring can serve as a lightweight and interpretable prioritization backbone, while text representations can help surface cases whose risk signals are expressed through vulnerability descriptions, product context, advisory language, or public-exploit text rather than compact tabular features alone. \linebreak \textsc{ThreatLens} should therefore be viewed as a triage compression layer, not an autonomous remediation oracle. Its goal is to reduce a large weekly CVE pool into a smaller, evidence-supported review set. The early-warning results show that public evidence often accumulates before KEV cataloging; \textsc{ThreatLens} identifies CVEs whose cutoff-valid evidence profile warrants earlier analyst attention. Downstream retrieval, explanation, and validation modules can then support investigation and remediation decisions.

The early-versus-late KEV diagnostics also caution against treating KEV timing as a single-score phenomenon. KEV entry reflects exploitation, disclosure, cataloging, and operational prioritization dynamics. In organizational deployments, the same CVE-time snapshot formulation could be extended with local exposure, asset criticality, patch status, and network reachability~\cite{nist80040r4}. This would shift the framework from global exploitation relevance toward organization-specific operational risk.

To illustrate this operational value, we discuss two held-out KEV CVEs from the forward-time early-warning evaluation. In both cases, \textsc{TL-LR} ranked the CVE within the top 10 before KEV inclusion, while EPSS never ranked it within the top 50 before KEV entry. For CVE-2023-52163 \cite{cve52163}, a command-injection vulnerability affecting DigiEver DS-2105 Pro network video recorders, \textsc{ThreatLens} ranked the CVE within the top 10 on 10 November 2025, 42 days before KEV inclusion; EPSS ranked it 152nd at the same cutoff. Fortinet later linked exploitation to the ShadowV2 IoT botnet, which targeted vulnerable devices for malware deployment and persistence~\cite{shadowv2}. This case shows that \textsc{ThreatLens} can prioritize operationally significant IoT vulnerabilities before they receive high EPSS rank or KEV confirmation.

For CVE-2026-20963 \cite{cve20963}, a Microsoft SharePoint remote-code-execution vulnerability, \textsc{ThreatLens} ranked the CVE 9th on 16 February 2026, 30 days before KEV inclusion; EPSS ranked it 166th at the same cutoff and never placed it within the top 50 before KEV entry. Government advisories later warned of active exploitation against internet-exposed SharePoint deployments and emphasized the risk of full server compromise~\cite{sharepoint2026}. This case highlights \textsc{ThreatLens}' ability to surface high-risk enterprise vulnerabilities before broader exploit-probability signals escalate.

The {\sc ThreatLens} results should be interpreted in light of several constraints.
\textsc{ThreatLens} uses KEV entry as weak supervision, but KEV is not a
complete exploitation record. It reflects confirmed exploitation,
remediation relevance, reporting practices, and cataloging decisions; thus,
some exploited CVEs may be treated as negatives, and some
organization-specific high-risk vulnerabilities may never enter KEV. Although
we mitigate leakage from NVD-derived fields by removing mutable metadata and
ablating CVSS/CWE/CPE-derived signals, we do not reconstruct exact historical
NVD revisions. Finally, \textsc{ThreatLens} estimates global exploitation
relevance rather than organization-specific risk. In deployment, its rankings
should be combined with local asset exposure, patch status, business
criticality, and network reachability.

\section{Conclusion}
\textsc{ThreatLens} shows that vulnerability prioritization can be treated as an operational early-warning problem rather than a static severity-ranking task. Its main contribution is a cutoff-aligned framework that continuously fuses public evidence about vulnerabilities, advisories, exploit availability, and exploitation likelihood to identify CVEs that deserve analyst attention before they are broadly recognized as urgent. In practical use, \textsc{ThreatLens} acts as a triage compression layer: it turns a large weekly stream of newly relevant CVEs into a smaller, evidence-backed review queue. This supports analysts who must make timely decisions under uncertainty, before all signals are complete or officially cataloged. Rather than replacing expert judgment, \textsc{ThreatLens} helps focus that judgment on vulnerabilities whose evolving evidence profile indicates emerging operational risk. More broadly, \textsc{ThreatLens} demonstrates that deployment-realistic vulnerability prioritization requires temporal discipline: models should be trained and evaluated using only evidence available at the time a decision would have been made. By combining temporally valid evidence fusion with analyst-oriented ranking, \textsc{ThreatLens} provides a practical foundation for early-warning vulnerability management and a stronger evaluation standard for future exploitation-prediction systems.

\newpage

\section{GenAI Usage Disclosure}

The authors used generative AI tools during multiple stages of the project. These tools were used to assist with writing and revision, including grammar, clarity, phrasing, organization, formatting, and improving the presentation of the manuscript. They were also used to support manuscript development and research communication, including brainstorming framing and terminology, checking consistency across sections, considering potential reviewer concerns, and refining explanations of experiments, results, tables, figures, and captions. Generative AI tools were additionally used as programming and data-processing assistance, including help with code review, debugging, implementation guidance, notebook/script organization, and checking the consistency of data-processing and evaluation logic.

All research design decisions, dataset construction, implementation work, experimental execution, analyses, numerical results, interpretation of results, claims, final manuscript content, and responsibility for the submitted work remain with the authors. The authors take full responsibility for the accuracy, originality, integrity, and presentation of the submitted work.

\bibliographystyle{ACM-Reference-Format}
\bibliography{references}

\end{document}